\documentclass{ws-ijmpe}
\usepackage{graphicx}
\usepackage{epstopdf}
\begin{document}
\title{Di-jet measurements in heavy-ion collisions at STAR}

\author{\footnotesize Elena Bruna for the STAR Collaboration}

\address{Physics Department, Yale University, 272 Whitney Avenue\\
 New Haven, CT-06511 USA\\elena.bruna@yale.edu}

\maketitle

\begin{history}
\received{(received date)}
\revised{(revised date)}
\end{history}

\begin{abstract}
Jets are produced from hard scatterings in the early stages of heavy-ion 
collisions. It is expected that these high-$p_T$ partons travel through the 
hot and dense medium before fragmenting.
Therefore they are expected to suffer energy loss in the QGP via gluon 
radiation and elastic collisions along their path.
Measurements of fully reconstructed jets help understand the effect of 
the energy loss on the jet structure and energy profile. A data-driven 
characterization of the background in Au+Au is needed in order to 
compare the results to p+p.
The full azimuthal coverage of STAR Time Projection Chamber
and Electromagnetic Calorimeter allows measurements of fully
reconstructed di-jets, defined by jets that match the online 
trigger and recoil jets on the away side. A tight selection of the 
trigger jets allows for a selection of those coming from the surface. 
Hence, the population of jets on the recoil side is biased towards a 
maximal energy loss because of the extreme in-medium pathlength.
We present measurements of di-jets, exploring their structure and 
properties in Au+Au and p+p at $\sqrt s_{NN}=$200 GeV in the STAR experiment.
\end{abstract}

\section{Introduction}
Jet quenching was first  inferred at RHIC via measurements of single hadron and di-hadron suppression at high $p_T$ in Au+Au collisions relative to p+p and d+Au~\cite{Raa}.
Although these observables showed evidence for parton energy loss in the medium, they were limited by the inability to constrain the kinematics of the hard-scattered parton and suffered from geometrical biases~\cite{renk}. 
Full jet reconstruction enables a more precise measurement of the energy of the initial hard scattering, reducing the biases in the study of jet quenching.
Di-jet measurements provide an additional tool to probe the hot QCD medium produced in high energy nuclear collisions.
By triggering on jets biased toward the surface of the medium, the corresponding recoil jets are exposed to a maximal path-length in the medium. Looking at recoil jets allows then the selection of a jet population that should suffer a large energy loss, in addition to the tangential emission.
Di-jet coincidence rates in p+p and Au+Au, as well as energy profile measurements of recoil jets are presented. Results of jet-hadron correlations requiring a reconstructed di-jet are also reported as a tool to assess  biases in jet finding, and to help understanding the quenching scenario.
The results are obtained utilizing modern jet-finding algorithms and data-driven correction schemes.

\section{Data sets and analysis}\label{ana}
The sub-detectors used for jet reconstruction in STAR are the Time Projection Chamber (TPC) for charged particles and the Barrel Electromagnetic Calorimeter (BEMC) for the neutral energy. Both TPC and BEMC have full azimuthal coverage and pseudo-rapidity acceptance $|\eta|<1$, making STAR a suitable detector for di-jet measurements.
Corrections for double-counting of electrons and charged hadronic energy deposition in the BEMC are applied. 
This analysis is based on p+p year 2006 and 0-20$\%$ most central Au+Au year 2007 events. Both data sets were selected with an online High-Tower (HT) trigger in the BEMC which requires the transverse energy in a tower to be  $E_T>5.4$ GeV.
The jet finding algorithms utilized are ``anti-$k_T$'' to reconstruct the jets and   ``$k_T$'' to measure the mean background per unit area. Both ``anti-$k_T$''  and  ``$k_T$'' are recombination algorithms included in the FastJet package~\cite{fastjet}. 
The di-jets are made of a  ``trigger'' jet, required to match the online triggered tower in the BEMC, and a ``recoil'' jet, reconstructed on the away side of the trigger jet (i.e. $|\Delta \phi| > 2.74$).
FastJet provides an estimate of the background $p_T$ per unit area, that is subtracted to get the reconstructed jet p$_{T}^{jet}$. Fluctuations around this value are approximated by a Gaussian distribution with $\sigma=6.5$ GeV and 3.7 GeV for jets reconstructed with resolution parameter R=0.4 and R=0.2 respectively and $p_T^{cut}=0.2$ GeV/c on tracks and neutral towers . The recoil spectra with $p_T^{cut}=2$ GeV/c are unfolded using $\sigma=1.5$ GeV for R=0.4. Work is in progress~\cite{mateusz}  to derive a functional form of the background fluctuations that accounts for the non-Gaussian tails.
The trigger jets were reconstructed with a resolution parameter R=0.4 and with a $p_T^{cut}=2$ GeV/c in order to have similar jet energy scale in p+p and Au+Au. This allows the selection of hard fragmenting jets produced in proximity of the surface of the medium, and/or less interacting. The requirement of $p_{T}^{jet}>20$ GeV/c for the reconstructed trigger jets minimizes the contamination by background jets. In order to study the bias introduced by a $p_T^{cut}$ on the recoil side, the recoil jets are reconstructed with two different choices of  $p_T^{cut}$, 0.2 and 2 GeV/c. 
Two choices of R were made for the recoil jets, R=0.4 and R=0.2, allowing energy profile measurements.
The recoil jet spectrum is corrected for the background jet spectrum, estimated from the $p_T$ distribution of associated jets located in a window centered at $\pm  \pi/2$ with respect to the trigger jet axis. 
The reconstructed jet momenta are corrected for the different tracking efficiencies in p+p and Au+Au.

\section{Di-jet coincidence rate and jet energy profile}\label{dijet}
\begin{figure}
\centering

\resizebox{0.47\textwidth}{!}{  \includegraphics{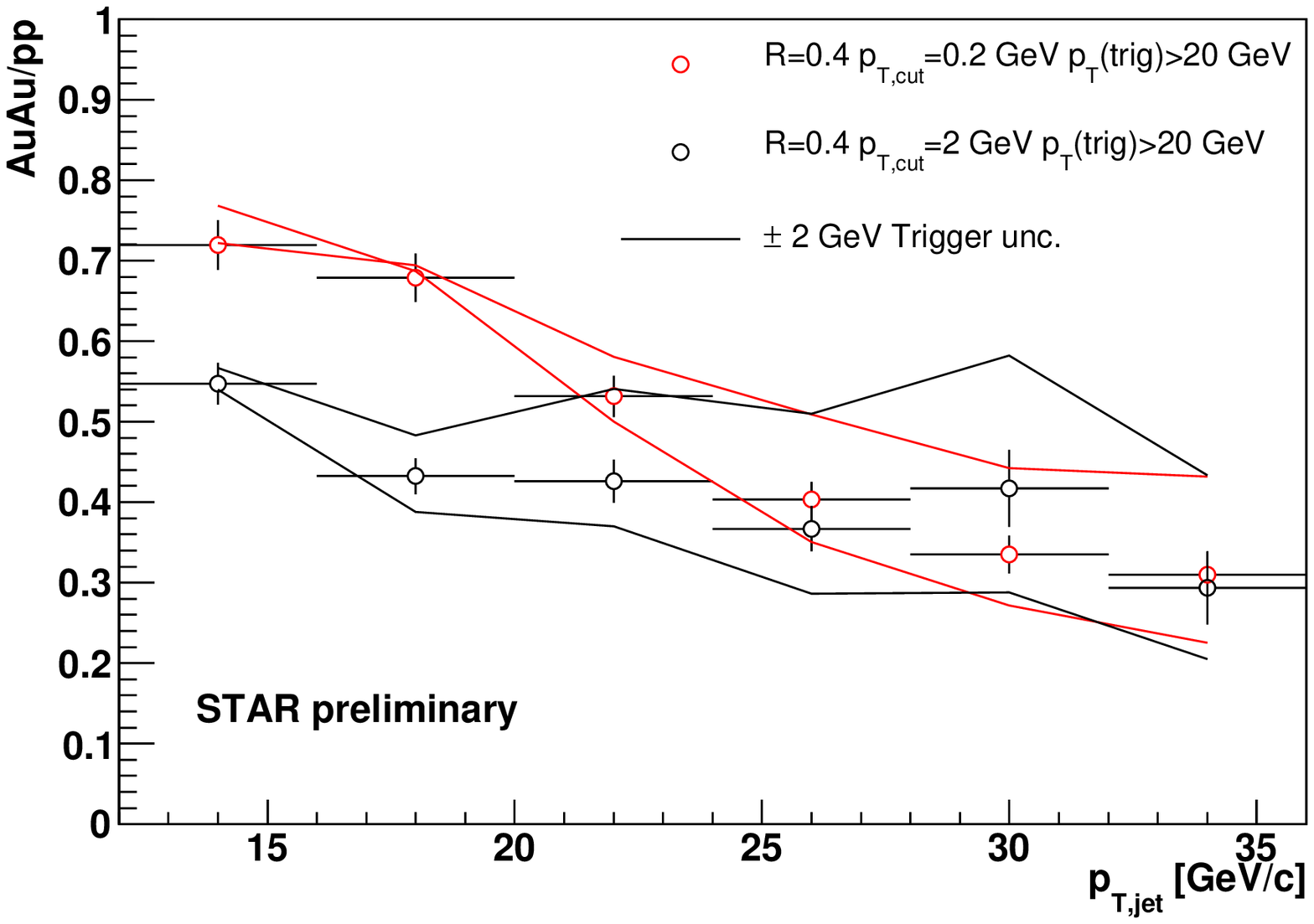}}
\resizebox{0.47\textwidth}{!}{  \includegraphics{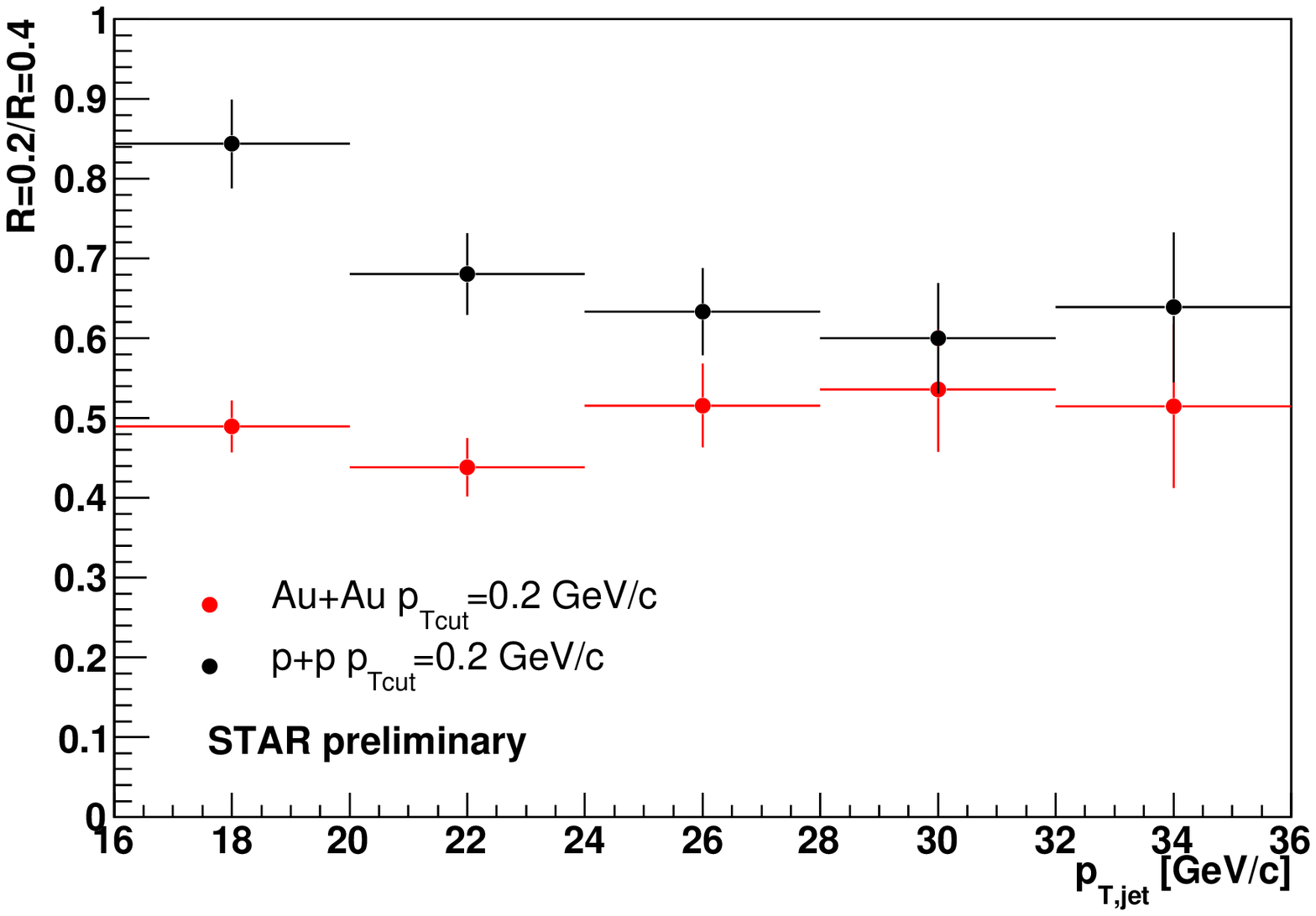}}

\caption{Left: Ratio ofÊ di-jet coincidence rates in Au+Au relative to p+p for two values of $p_T^{cut}$ on the recoil side, 0.2 and 2 GeV/c. The solid curves represent the uncertainty on the trigger jet energy. Right: Ratio of di-jet coincidence rates R=0.2/R=0.4 for recoil jets in p+p and Au+Au. The results are obtained with  p$_T^{trig}>20$ GeV/c. }
\label{fig:dijet}

\end {figure}

The ratio of di-jet spectra of Au+Au relative to p+p is reported in Fig.~\ref{fig:dijet} (left). 
A similar suppression is observed for the two different values of $p_T^{cut}$. 
A main requirement in this analysis is the control of the trigger energy in the two different collision systems. 
Even with  a tight $p_T^{cut}$ in Au+Au, the reconstructed jet momentum might be affected by upward background fluctuations of the order of 1.5 GeV, that artificially enhance the measured $p_{T,jet}$ compared to the same jet in p+p. On the other hand, given that trigger jets in Au+Au may actually interact in the medium~\cite{joern}, the measured $p_{T,jet}$ could correspond to a higher momentum jet in p+p. Both these effects are quantified by varying the trigger jet momentum in p+p by $\pm 2$ GeV/c, as indicated by the solid curves in  Fig.~\ref{fig:dijet} (left).
The observed suppression can be explained via a broadening of the jet energy profile in the recoil side, due to energy loss to large angles. 
One way to assess this effect is to study the energy profile for two choices of the resolution parameter R. Results from inclusive jet analysis indicate substantial broadening from R=0.2 to R=0.4 in Au+Au central collisions~\cite{mateusz}.
Fig.~\ref{fig:dijet} (right) shows the ratio of di-jet coincidence rates R=0.2/R=0.4 for recoil jets in p+p and Au+Au. The Au+Au ratio is lower than p+p but it is less suppressed compared to the inclusive analysis, because of the flatter recoil spectrum.
Assuming a simple picture where the broadening is responsible for a constant shift of the Au+Au recoil spectrum towards lower p$_T$ relative to p+p, we estimated that a shift of $8-9$ GeV/c for R=0.4 and of $7-8$ GeV for R=0.2 accommodates the measured ratios in Fig.~\ref{fig:dijet}. 

\section{Results from jet-hadron correlations with reconstructed di-jets}
\begin{figure}
\centering

\resizebox{0.47\textwidth}{!}{  \includegraphics{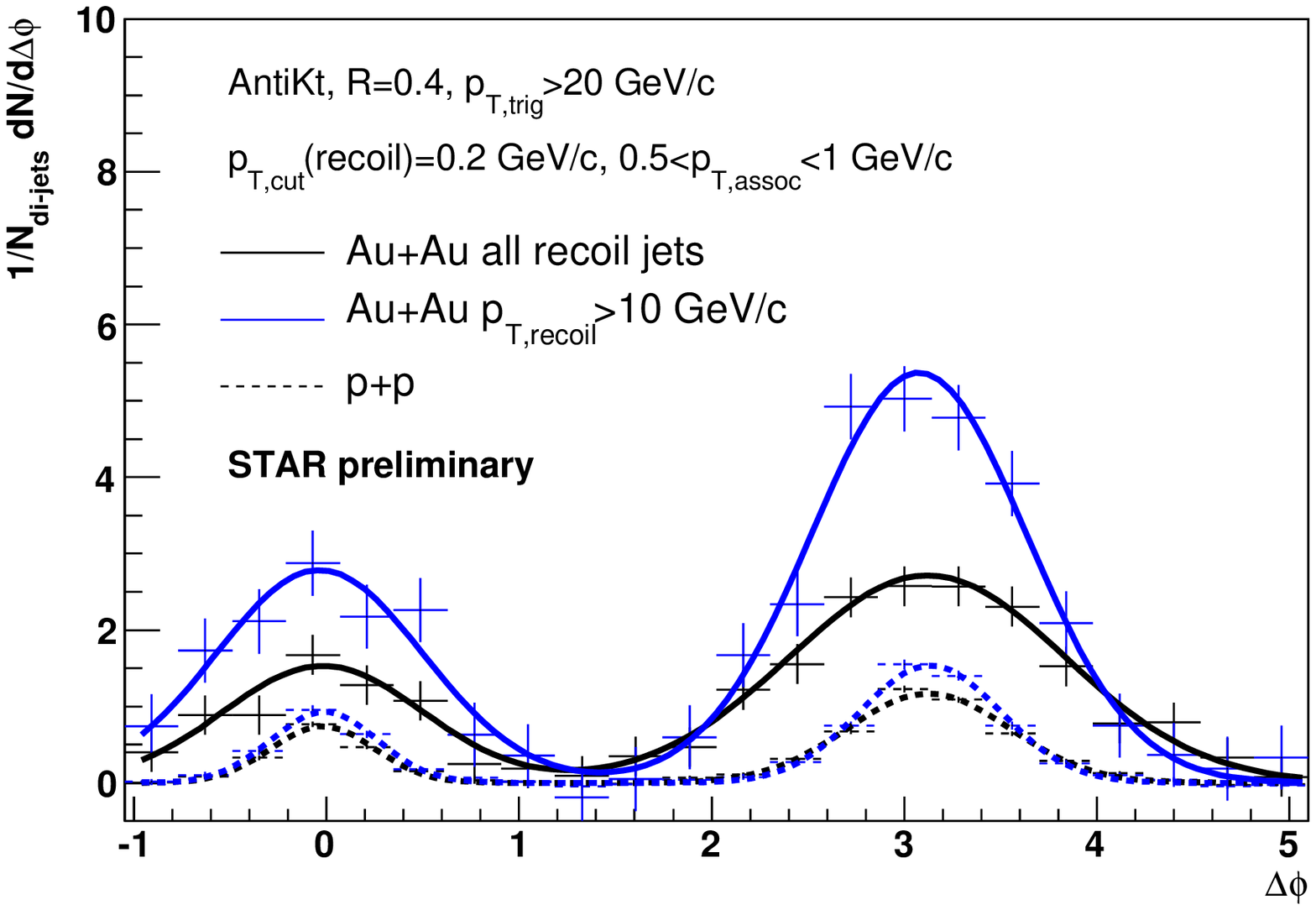}}
\resizebox{0.47\textwidth}{!}{  \includegraphics{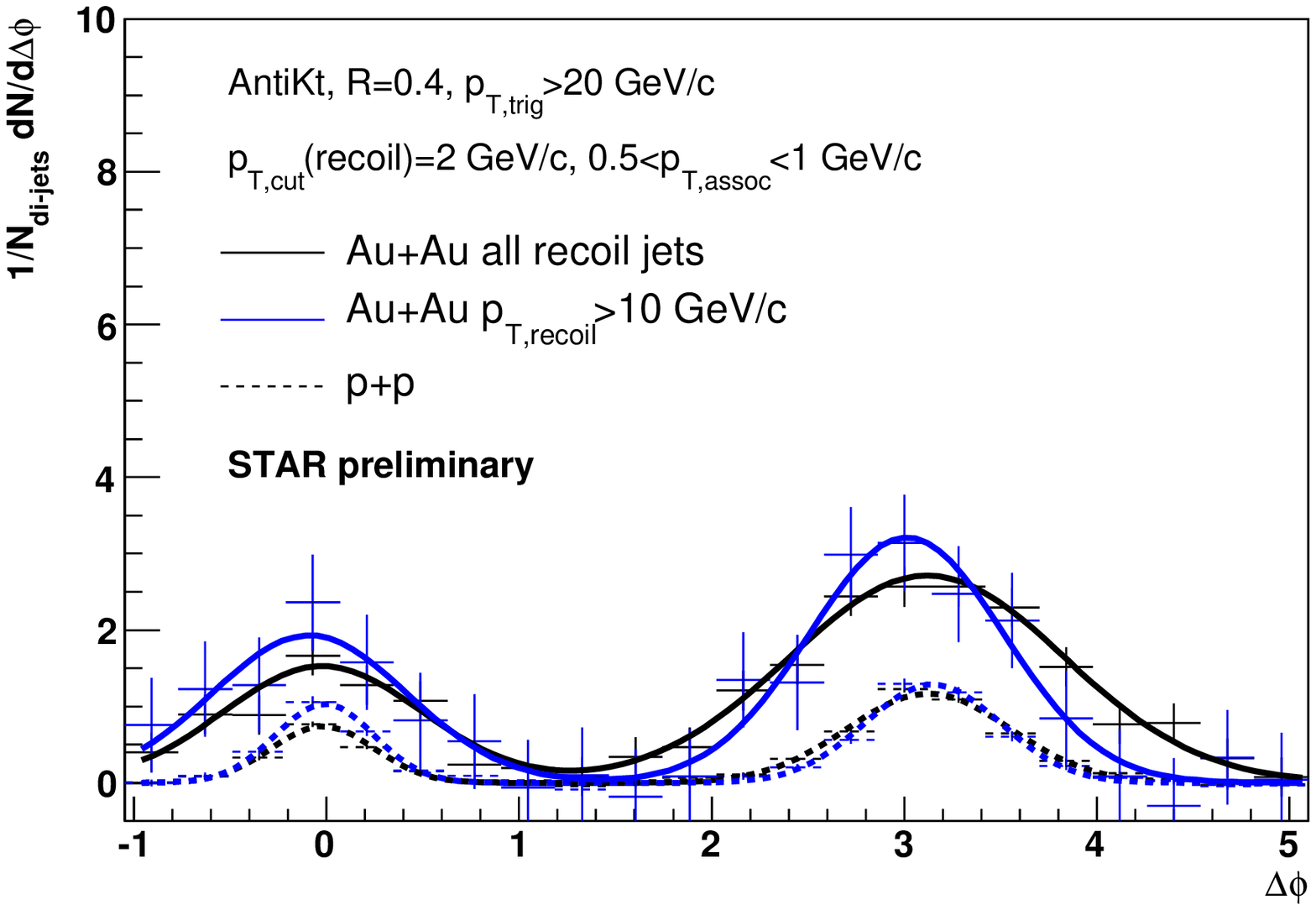}}
\caption{Left: Jet-hadron correlations for $0.5<p_{T,assoc}<1$ GeV/c in Au+Au (solid lines) and p+p (dashed), for all events with recoil jets (black) and for events with recoil jets selected above $p_{T,recoil}>10$ GeV/c (blue). Recoil jets are reconstructed  with  $p_T^{cut}=0.2$ GeV/c. Right: same as left but $p_T^{cut}=2$ GeV/c is used to reconstruct the recoil jets.}
\label{fig:jh1}

\end {figure}

\begin{figure}
\centering

\resizebox{0.47\textwidth}{!}{  \includegraphics{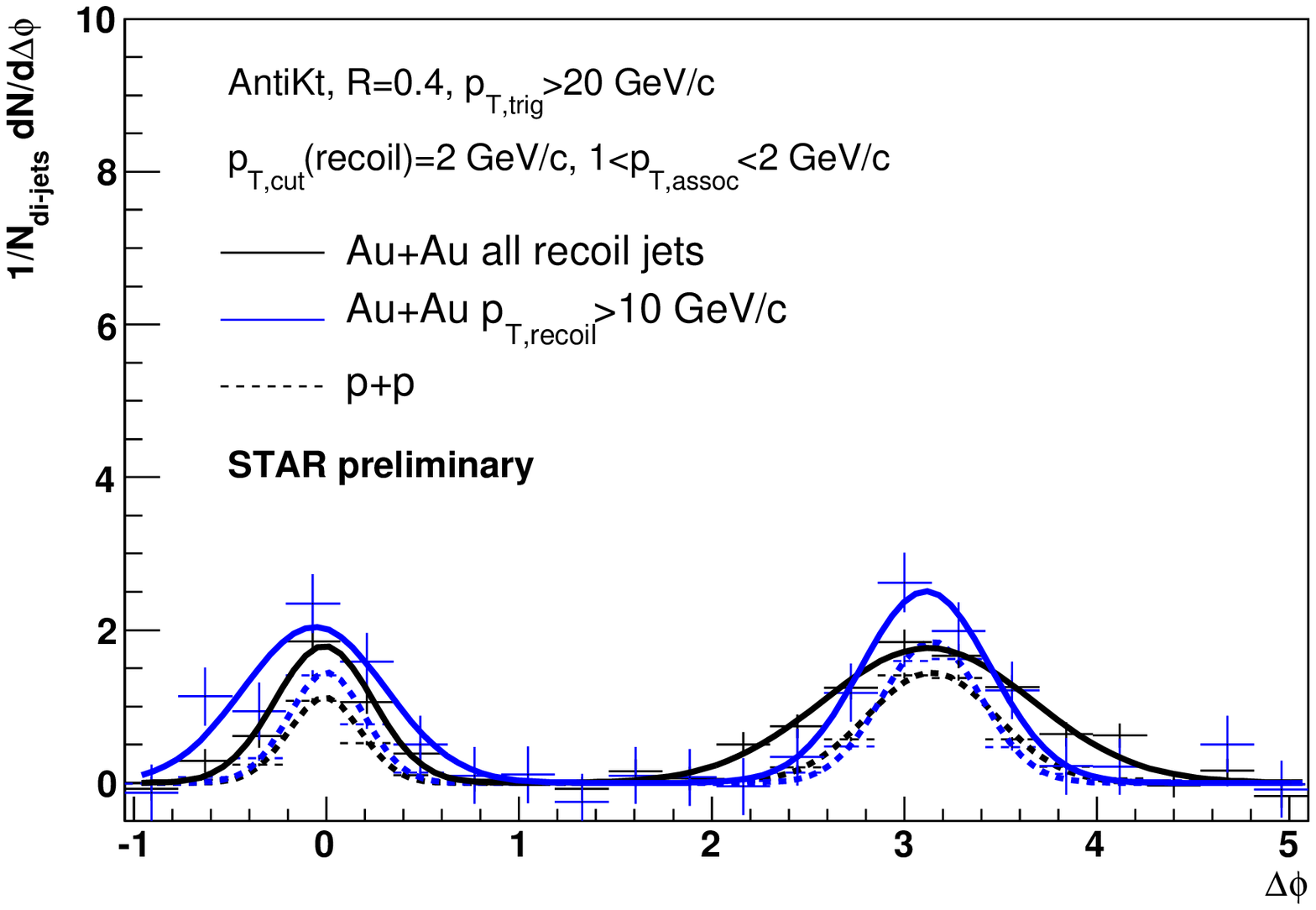}}
\resizebox{0.47\textwidth}{!}{  \includegraphics{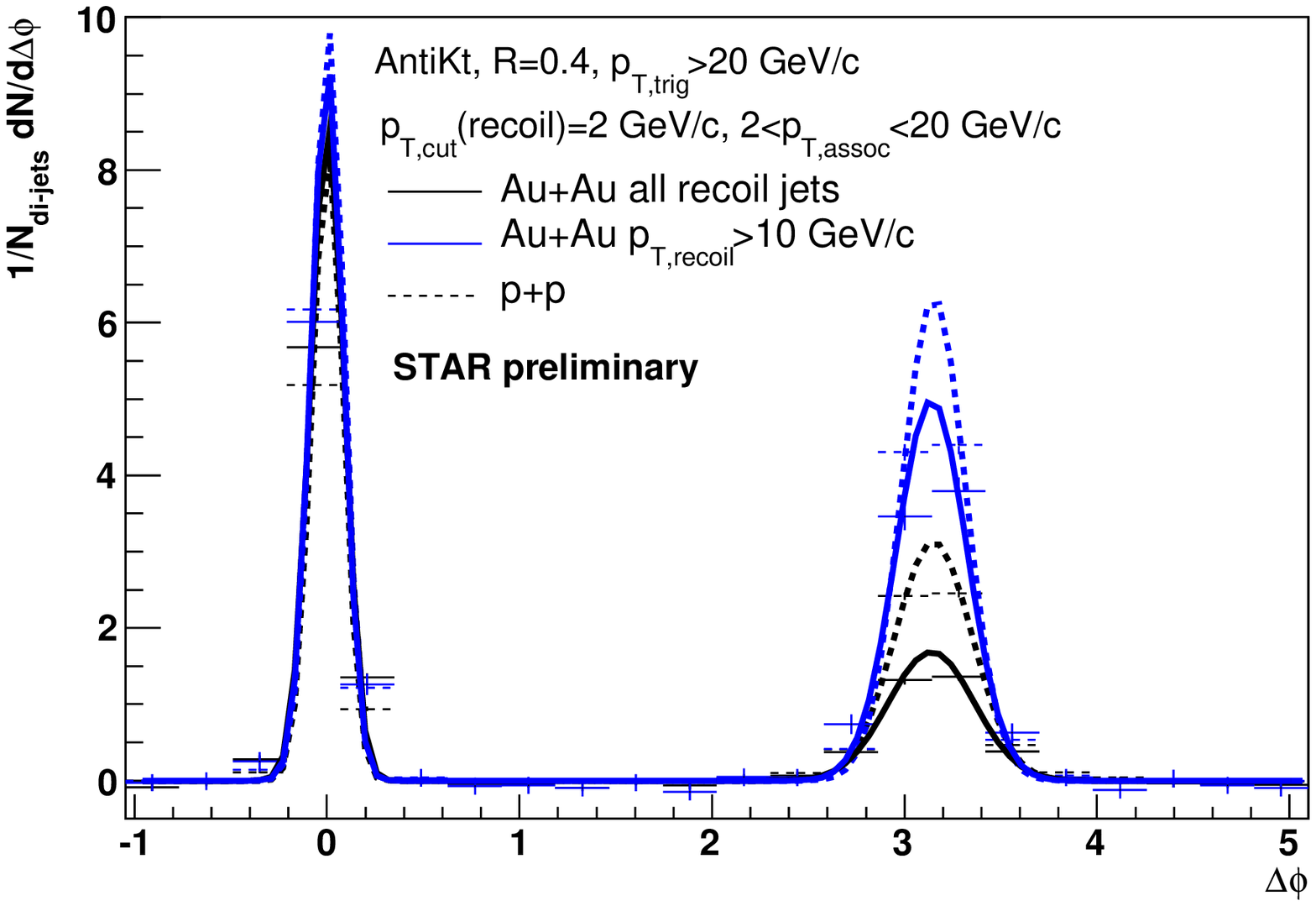}}
\caption{Left: Jet-hadron correlations for $1<p_{T,assoc}<2$ GeV/c in Au+Au (solid lines) and p+p (dashed), for all events with recoil jets (black) and for events with recoil jets selected above $p_{T,recoil}>10$ GeV/c (blue). Recoil jets are reconstructed  with  $p_T^{cut}=2$ GeV/c. Right: same as left but $2<p_{T,assoc}<20$ GeV/c is used in the correlation.}
\label{fig:jh2}

\end {figure}

Results reported above are in qualitative agreement with a scenario of jet broadening outside of R=0.4. 
Recoil jets reconstructed with $p_T^{cut}=2$ GeV/c are also significantly suppressed. 
The question we want to address is whether such jets are less interacting, as we would expect because of the bias given by the  $p_T^{cut}$. 
Jet-hadron correlations are a powerful tool to investigate the quenching mechanism. Preliminary results on jet-hadron correlations are in agreement with a broadening scenario and softening of the jet fragmentation~\cite{joern}.
This analysis is a first look at correlations of hadrons with respect to the trigger jets in events where di-jets are reconstructed. This procedure aids the understanding of possible biases in the jet-finding via a direct measurement of the recoil side structure. Trigger and recoil jets are selected as described in Sec.~\ref{ana}, both in p+p and Au+Au. 
Fig.~\ref{fig:jh1} shows the azimuthal jet-hadron correlations for  $0.5<p_{T,assoc}<1$ GeV/c in Au+Au (solid lines) and p+p (dashed), for all recoil jets (black) and for recoil jets selected above $p_{T,recoil}>10$ GeV/c (blue), reconstructed with $p_T^{cut}=0.2$ GeV/c (left) and $p_T^{cut}=2$ GeV/c (right). 
The results are not corrected for fake jets. However, this correction will enhance the away side in Au+Au with respect to p+p and is expected to be small with  $p_T^{cut}=2$ GeV/c, where the background jets are about $1/30$ of the recoil jets reconstructed with $p_{T,jet}>10$ GeV/c.
For $p_T^{cut}=0.2$ GeV/c about $1/3$ of the recoil jets reconstructed with $p_{T,jet}>10$ GeV/c are expected to be background jets.
To avoid possible fake correlations due to background jets made of low-$p_T$ randomly clustered particles, the jet-hadron correlations are performed with associated hadron $p_T$ above 0.5 GeV/c.
A qualitative comparison of the away side in Au+Au relative to p+p  shows an enhancement of low-$p_T$ particles on the away side, as well as a broadening. 
This effect is emphasized with the presence of a recoil jet above $p_{T,recoil}>10$ GeV/c, where the probability of hadrons coming from di-jet fragmentation and hence correlated with the trigger jet is higher.
The away side yield in Au+Au relative to p+p is lower when recoil jets are reconstructed with $p_T^{cut}=2$ GeV/c compared to $p_T^{cut}=0.2$ GeV/c, suggesting that those jets are less interacting, as expected because of the bias in the jet-finding.
At higher  $p_{T,assoc}$, as reported in Fig.~\ref{fig:jh2} for $1<p_{T,assoc}<2$ GeV/c (left) and $2<p_{T,assoc}<20$ GeV/c (right), a decrease of high-$p_T$ associated particles on the away side is clearly visible in Au+Au with respect to p+p for $p_T^{cut}=2$ GeV/c, suggesting that also recoil jets reconstructed with a larger bias are subject to the effects of energy loss.
 
\section{Summary}
Di-jet measurements were presented in these proceedings. The suppression of the di-jet coincidence rate in Au+Au relative to p+p, as well as the ratios of jet spectra with different R parameters in Au+Au and p+p, are in qualitative agreement with a scenario of jet broadening outside of R=0.4. This observation can be interpreted as the effect of the long in-medium path that a large population of recoil jets (i.e. those non-tangentially emitted) is expected to be exposed to, because of the tight requirement on the trigger jets.
The measurements of jet-hadron correlations in the presence of reconstructed di-jets are used to address possible biases in the jet-finding. In particular, different values of $p_{T,cut}$ on the recoil jets help in understanding the jet quenching mechanism. These first results indicate that even highly biased di-jets reconstructed  with  $p_T^{cut}=2$ GeV/c are interacting, supporting the hypothesis of a broadening scenario rather than full absorption in the medium.
Full set of corrections, including fake jets, as well a more complete assessment  of the background subtraction and of the trigger bias are ongoing in order to have a quantitative estimate of the out-of-cone energy loss via jet-hadron correlations in presence of recoil jets.
The use of models will be helpful to compare the measured out-of-cone energy with the theoretical predictions.

\section{Acknowledgments}
The author wishes to thank the Bulldog computing facility at the Yale University.


\begin{thebibliography}{0}
\bibitem{Raa} J Adams et al., STAR Collaboration,  Nucl. Phys. A 757 (2005) 102.
\bibitem{renk} T. Renk, K. Eskola, Phys. Rev. C75 (2007) 054910, arXiv:hep-ph/0610059.
\bibitem{fastjet} M. Cacciari, G. Salam, G. Soyez, arXiv:0802.1188.
\bibitem{mateusz} M. Ploskon for the STAR Collaboration, these proceedings.
\bibitem{joern} J. Putschke for the STAR Collaboration, these proceedings.
\end{thebibliography}
\end{document}